\documentstyle[twoside,11pt,epsfig,color]{article}
\begin{document}

\title{Scalar field collapse in a conformally flat spacetime}

\author{Soumya Chakrabarti\footnote{email: adhpagla@iiserkol.ac.in}~~~and~~Narayan Banerjee\footnote{email: narayan@iiserkol.ac.in}
 \\
Department of Physical Sciences, \\Indian Institute of Science Education and Research, Kolkata\\ Mohanpur Campus, West Bengal 741252, India.
\date{}
}

\maketitle
\vspace{0.5cm}
{\em PACS Nos. 04.50.Kd; 04.70.Bw
\par Keywords : gravitational collapse, spherical symmetry, conformally flat, scalar field, singularity, horizon}
\vspace{0.5cm}

\pagestyle{myheadings}
\newcommand{\be}{\begin{equation}}
\newcommand{\ee}{\end{equation}}
\newcommand{\bea}{\begin{eqnarray}}
\newcommand{\eea}{\end{eqnarray}}

\begin{abstract}
The collapse scenario of a scalar field along with a perfect fluid distribution is investigated for a conformally flat spacetime. The theorem for the integrability of an anharmonic oscillator has been utilized. For a pure power law potential of the form ${\phi}^{n+1}$, it is found that a central singularity is formed which is covered by an apparent horizon for $n>0$ and $n<-3$. Some numerical results have also been presented for a combination of two different powers of $\phi$ in the potential.
\end{abstract}

\section{Introduction}

The final outcome of a continuous gravitational collapse has been of great interest in general relativity. During the final  stages of a stellar evolution, one is left only with gravitational interaction, which being attractive, a collapse is always on the cards. The natural question to ask is about the end product of this collapse. The first systematic analysis of an unhindered gravitational collapse in general relativity was given by Oppenheimer and Snyder \cite{Opp}. The general belief is  that the ultimate spacetime singularity, which might be hit by the collapsing matter, is actually shielded from an exterior observer by the formation of an event horizon. However, this idea suffers a jolt by the explicit examples of collapsing models, with perfectly reasonable matter distribution, which show the absence of an event horizon, leading to what is known as a naked singularity. For a comprehensive systematic description of the gravitational collapse, we refer to the monograph by Joshi \cite{pankaj1}. More recent work and various aspects of collapse have been systematically summarized in a recent work by Joshi \cite{pankaj2}. \\
Scalar fields, albeit having no pressing motivation from particle physics theory more often than not, have been of great interest in theories of gravity for various reasons. A scalar field with a variety of potential can mimic the evolution of many a kind of matter distribution. For instance, Goncalves and Moss \cite{gong2} showed that the collapse of a spherically symmetric scalar field can be formally treated as a collapsing dust ball. A scalar field fits in superbly for cosmological requirements such as the driver of the past or even  the present acceleration of the universe. Various massive scalar fields have gained a lot of interest during the past two decades for their role as the driver of the recent accelerated expansion of the universe. The collapse scenario of such fields have also been studied recently \cite{gong2, giambo, gong1, goswami2, koyel, cai1, cai2}.              \\
A zero mass scalar field collapse was discussed by Christodoulou \cite{christo1}. Christodoulou also showed the possibility of the formation of a naked singularity as an end product of a scalar field collapse \cite{christo3}. With the help of a numerical analysis, Goldwirth and Piran, however, showed that a scalar collapse leads to a singularity which is cut-off from the exterior observer by an event horizon \cite{piran}. Some extensive investigations are already there in the literature in connection with a scalar field collapse, particularly in a spherically symmetric spacetime. These investigations indicate a very rich store of  possibilities. For example, some numerical calculations, by Choptuik \cite{chop}, Brady \cite{brady} and Gundlach \cite{gund} indicate that a scalar field collapse may lead to some critical phenomena close to the threshold of black hole horizon. For a comprehensive review of the critical phenomena associated with a scalar field collapse, we refer to the work of Gundlach \cite{Gundlach1}. Scalar field collapse have been analytically studied by Goswami and Joshi \cite{goswami}, Giambo \cite{giambo} quite recently. \\

The aim of the present work is to look at the collapse of a massive scalar field along with a distribution of perfect fluid. The potential is taken to be a power law ($V\sim {\phi}^{n+1}$) where $n$ can take a wide range of values. Indeed the system of equations is notoriously nonlinear and thus discussions on such fields, particularly analytical studies, are limited to very special cases. We start with a conformally flat spacetime. Recently Sharma, Das and Tikekar \cite{ranjan} used such a spacetime for the study of collapse of a fluid with a heat flux. We investigate the scalar field analogue of their work without a heat flux.\\ 

Even with an apparently simple metric, it is extremely difficult to handle the system of equations analytically. We adopt a completely different strategy. The integrability conditions for an anharmonic oscillator, developed by Euler \cite{euler} and utilized by Harko, Lobo and Mak \cite{harko} for a power law potential is invoked. This leads to an integrable second order equation for the scale factor. Thus some general comments, regarding the possibility and nature of a wide range of power law potentials can be made.\\

Most of the investigations regarding collapse with a scalar field deals with situations where there is no contribution from any fluid in the matter sector. The present work deals with the scalar field collapse along with a fluid distribution. No equation of state for the fluid is assumed at the outset. The relevance of such investigations stems from the present importance of a scalar field as the dark energy \cite{paddy, varun, sami}, the agent responsible for the late time acceleration of the universe. It also deserves mention that the distribution of the dark energy {\it vis-a-vis} the fluid is not known. It is generally believed that the dark energy does not cluster at any scale below the Hubble scale. The study of collapse of scalar fields, particularly in the presence of a fluid may in some way enlighten us regarding the possible clustering of dark energy. \\

The paper is organized as follows. In the second section we write down the relevant equations for a scalar field model along with a perfect fluid distribution in a conformally flat spherically symmetric spacetime. The third section includes a very brief review of the integrability of anharmonic oscillator equation following the work of Euler \cite{euler} and Harko {\it et al} \cite{harko}. In the fourth section, we investigate the collapse of a scalar field with a power-law potential in detail. The discussion on a combination of a quadratic  and an arbitrary power-law potential is given in section five. The sixth and the final section includes a summary of the results obtained.

\section{Conformally flat metric and a scalar field collapse}
We write the metric for a spherically symmetric spacetime as 
\begin{equation}
\label{metric}
ds^2=\frac{1}{{A(r,t)}^2}\Bigg[dt^2-\frac{dr^2}{1-kr^2}-r^2d\Omega^2\Bigg].
\end{equation}

The time evolution is governed solely by the function A(r,t). The Weyl tensor for the spacetime metric vanishes, thus this metric admits a  conformal flatness. \\

The energy momentum tensor for a perfect fluid is given by
\begin{equation}
\label{em-tensor}
T^{m}_{\mu\nu}=(\rho+p)u_{\mu}u_{\nu}-p g_{\mu\nu},
\end{equation}

where $\rho$ is the energy density, $p$ is the isotropic fluid pressure, $u_{\mu}$ is the 4-velocity of the fluid.      \\
When a scalar field $\phi$ is minimally coupled to gravity, the relevant action is given by 
\begin{equation}
\textit{A}=\int{\sqrt{-g}d^4x[R+\frac{1}{2}\partial^{\mu}\phi\partial_{\nu}\phi-V(\phi) + L_{m}]},
\end{equation}

where $V(\phi)$ is the self-interaction potential of the scalar field and $L_{m}$ is the Lagrangian density for the fluid distribution. From this action, the contribution to the energy-momentum tensor from the scalar field $\phi$ can be written as
\begin{equation}
T^\phi_{\mu\nu}=\partial_\mu\phi\partial_\nu\phi-g_{\mu\nu}\Bigg[\frac{1}{2}g^{\alpha\beta}\partial_\alpha\phi\partial_\beta\phi-V(\phi)\Bigg]. 
\end{equation}

Einstein field equations $G_{\mu\nu} = - T_{\mu\nu}$ (in the units $8 \pi G = 1$) can thus be written as
\begin{equation}
\label{fe1}
3kA^2+3\dot{A}^2-3(1-kr^2)A'^2+2(1-kr^2)AA''+\frac{2(2-3kr^2)}{r}AA'=\rho+\frac{1}{2}A^2\dot{\phi}^2-\frac{1}{2}A^2(1-kr^2)\phi'^2+V(\phi),
\end{equation}

\begin{equation}
\label{fe2}
-kA^2+2\ddot{A}A-3\dot{A}^2+3(1-kr^2)A'^2-\frac{4}{r}(1-kr^2)AA'=p+\frac{1}{2}{\phi'}^2A^2(1-kr^2)+\frac{1}{2}A^2\dot{\phi}^2-V(\phi),
\end{equation}

\begin{equation}
\label{fe3}
-kA^2+2\ddot{A}A-3\dot{A}^2+3(1-kr^2)A'^2-\frac{2}{r}(1-2kr^2)AA'-2(1-kr^2)AA''=p-\frac{1}{2}{\phi'}^2A^2(1-kr^2)+\frac{1}{2}A^2\dot{\phi}^2-V(\phi),
\end{equation}

\begin{equation}
\label{fe4}
\frac{2\dot{A}'}{A}=\dot{\phi}\phi' .
\end{equation}

\par The wave equation for the scalar field is given by
\begin{equation}
\label{wave}
\Box\phi+\frac{dV}{d\phi}=0.
\end{equation}

For the sake of simpliciy, we assume that $\phi(r,t)$ is a function of time $t$ alone. Consequently, from equation (\ref{fe4}), one can see that $A(r,t)$ can also be written as a function of time alone, and this is consistent with the rest of the equations of the system. With this, equations (\ref{fe2}) and (\ref{fe3}) become identical. So effectively we shall be considering a collapse of a perfect fluid and scalar field with a spatial homogeneity, analogous to the Oppenheimer-Snyder collapse of a spatially homogeneous fluid \cite{Opp}. \\

With $A=A(t)$ and $\phi=\phi(t)$, the field equations simplify as

\begin{equation}
\label{fenew1}
3kA^2+3\dot{A}^2=\rho+\frac{1}{2}A^2\dot{\phi}^2+V(\phi),
\end{equation}

\begin{equation}
\label{fenew2}
-kA^2+2\ddot{A}A-3\dot{A}^2=p+\frac{1}{2}A^2\dot{\phi}^2-V(\phi),
\end{equation}

and the wave equation looks like
\begin{equation}
\label{wave2}
\ddot{\phi}-2\frac{\dot{A}}{A}\dot{\phi}+\frac{1}{A^2}\frac{dV}{d\phi}=0.
\end{equation}

Now we have three equations (\ref{fenew1} - \ref{wave2}) to solve for four unknowns, namely $A, \phi, \rho$ and $p$. $V$ of course is given as a function of $\phi$. Instead of choosing an equation of state given as $p = p(\rho)$ in order to close the system, we shall utilize the condition for integrability of the equation (\ref{wave2}) to get the solution for the scale factor.

\section{A note on the integrability of anharmonic oscillator equation}
A nonlinear anharmonic oscillator with variable coefficients and a power law potential can be written in a general form as

\begin{equation}
\label{gen}
\ddot{\phi}+f_1(t)\dot{\phi}+ f_2(t)\phi+f_3(t)\phi^n=0,
\end{equation}

where $f_i$-s are functions of $t$, and $n \in {\cal Q}$ is a constant. An overhead dot represents a differentiation with respect to time $t$. Using Euler’s theorem on the integrability of the general anharmonic oscillator equation \cite{euler} and recent results given  by Harko {\it et al} \cite{harko}, this equation can be integrated under certain conditions. The essence can be written in the form of a theorem as discussed in \cite{euler, harko}.

{\bf \large Theorem} An equation of the form of equation (\ref{gen}) can be transformed into an integrable form 
\begin{equation}
\label{Phi1}
\frac{d^{2}\Phi}{dT^{2}}+\Phi^{n}\left( T\right) =0.
\end{equation}
by introducing a pair of new variables $\Phi$ and $T$ given by 
\begin{eqnarray}
\label{Phi}
\Phi\left( T\right) &=&C\phi\left( t\right) f_{3}^{\frac{1}{n+3}}\left( t\right)
e^{\frac{2}{n+3}\int^{t}f_{1}\left( x \right) dx },\\
\label{T}
T\left( \phi,t\right) &=&C^{\frac{1-n}{2}}\int^{t}f_{3}^{\frac{2}{n+3}}\left(
\xi \right) e^{\left( \frac{1-n}{n+3}\right) \int^{\xi }f_{1}\left( x
\right) dx }d\xi,
\end{eqnarray}%
if and only if $n\notin \left\{-3,-1,0,1\right\} $, and the coefficients of Eq. (\ref{gen}) satisfy the differential condition
\begin{equation}
\label{int-gen}
\frac{1}{n+3}\frac{1}{f_{3}(t)}\frac{d^{2}f_{3}}{dt^{2}%
}-\frac{n+4}{\left( n+3\right) ^{2}}\left[ \frac{1}{f_{3}(t)}\frac{df_{3}}{dt%
}\right] ^{2}+ \frac{n-1}{\left( n+3\right) ^{2}}\left[ \frac{1}{f_{3}(t)}%
\frac{df_{3}}{dt}\right] f_{1}\left( t\right) + \frac{2}{n+3}\frac{df_{1}}{dt}%
+\frac{2\left( n+1\right) }{\left( n+3\right) ^{2}}f_{1}^{2}\left( t\right)=f_{2}(t).
\end{equation} 
where $C$ is a constant. \\

In what follows, we shall use this integrability condition in order to extract as much information as we can from the scalar field equation (\ref{wave2}) for some given forms of the potential $V=V(\phi)$.

\section{Power-law potential}
In the first example we assume that the potential is a power function of $\phi$, $V(\phi) = \frac{V_{0}{\phi}^{(n+1)}}{n+1} $ such that 
\begin{equation}
\label{power-law}
\frac{dV}{d\phi}= V_{0}\phi^n,
\end{equation}
where $n \in {\cal Q}$ and $V_{0}$ is a constant. While the potential with a postive power of $\phi$, where $\frac{d^{2}V}{d{\phi}^{2}}$ evaluated at $\phi = 0$ gives the mass of the field, is quite well addressed, potentials with inverse powers of $\phi$ are also quite useful in a cosmological context, particularly as tracking quintessence fields. Ratra and Peebles \cite{ratra} used a potential of the form $V = \frac{M^{4+\alpha}}{\phi^{\alpha}}$, where $M$ is the Planck mass. Similar potential had later been used as a tracker field by Steinhardt, Wang and Zlatev \cite{zlat} where $M$ loses the significance as the Planck mass and is rather used as a parameter to be fixed by observations.

\subsection{Integrability of the scalar field equation and the solution for the metric}
With this power law potential, the scalar field equation (\ref{wave2}) becomes

\begin{equation}
\label{wave-pwrlw}
\ddot{\phi}-2\frac{\dot{A}}{A}\dot{\phi}+\frac{V_{0}}{A^2}\phi^n=0,
\end{equation}

which can be written in a more general form of second order ordinary differential equation with variable coefficients as

\begin{equation}
\label{wave-pwrlw1}
\ddot{\phi}+f_1(t)\dot{\phi}+f_3(t)\phi^n=0,
\end{equation}

where $f_{i}(t)$-s are functions of time, determined by $A(t)$ and its derivatives. Equation (\ref{wave-pwrlw1}) is easily identified to be a special case of equation (\ref{gen}) with $f_1(t)=-2\frac{\dot{A}}{A}$, $f_{2} = 0$ and $f_3(t)=\frac{V_{0}}{A^2}$. Hence, the integrability condition as in equation (\ref{int-gen}) yields a second order differential equation of $A(t)$ in the form

\begin{equation}
\label{evolutionA}
-\frac{6}{(n+3)}\frac{\ddot{A}}{A}+\frac{18(n+1)}{(n+3)^2}\frac{\dot{A}^2}{A^2}=0.
\end{equation}

This can be integrated to yield an exact time-evolution of $A(t)$ as
\begin{equation}
\label{A}
A(t)=\Bigg[\frac{2n\sqrt{\lambda}}{(n+3)}(t_{0}-t)\Bigg]^{-(\frac{n+3}{2n})},
\end{equation}
where $\lambda$ is a constant of integration coming from the first integral and is a positive real number. It is interesting to note that the conformal factor is independent of the choice of $V_{0}$. \\

As the theorem is valid for $n\notin \left\{-3,-1,0,1\right\} $, we exclude these values of $n$ in the subsequent discussion. The radius of the two-sphere is given by $rY(t)$ where 
\begin{equation}
\label{Y}
Y(t) = \frac{1}{A(t)} =\Bigg[\frac{2n\sqrt{\lambda}}{(n+3)}(t_{0}-t)\Bigg]^{(\frac{n+3}{2n})}.
\end{equation} 
From (\ref{A}) and (\ref{Y}), the time evolution of the collapsing fluid can be discussed for different choices of the potential.
\begin{itemize}
\item {For $n > 0$ as well as for $n < -3$, one has $(\frac{n+3}{2n}) > 0$. Let us write $(\frac{n+3}{2n}) = {n_{0}}^2$. Then from equation (\ref{Y}) one can write the radius of the two-sphere as 
\begin{equation}\label{scale1}
rY(t) = r\Bigg[\frac{\sqrt{\lambda}}{{n_{0}}^2}(t_{0}-t)\Bigg]^{{n_{0}}^2}.
\end{equation} 
It is straightforward to note that $rY(t)$ goes to zero when $t \rightarrow t_0$. Thus, for all $n>0$ and for $n<-3$, the collapsing sphere reaches a singularity of zero proper volume at a finite time defined by $t_0$. We must exclude the case for $n = 1$ since this does not fall in the domain of the validity of the theorem.}

\item {However, for $0 > n > -3$, one has $(\frac{n+3}{2n}) < 0$ and it can be written as $(\frac{n+3}{2n}) = -{m_{0}}^2$. For this domain of $n$, the scale factor $Y$ can be written as
\begin{equation}\label{scale2}
rY(t) = r\Bigg[\frac{\sqrt{\lambda}}{{m_{0}}^2}(t-t_{0})\Bigg]^{-{m_{0}}^2}.
\end{equation}
Clearly, we have a collapsing solution as it is easy to check that $\dot{Y(t)} < 0$. However, the collapsing fluid reaches the zero proper volume only when $t \rightarrow \infty$. This indicates the system is collapsing forever rather than crushing to a zero proper volume singularity at a finite time. The case $n = -1$ falls outside the domain of validity of the integrability condition. In this case, the proper time $\tau$ is defined as $\frac{d\tau}{dt} = \Big[\frac{\sqrt{\lambda}}{m_{0}^2}(t-t_{0})\Big]^{\frac{m_{0}^2}{2}}$ which is positive. Therefore $\tau$ is a monotonically increasing function of $t$. Thus the conclusion is infact true against the proper time as well.}

\item {From (\ref{scale1}) and (\ref{scale2}), one can check that $\frac{dY(t)}{dt} < 0$ for all relevant cases, provided $\sqrt{\lambda} > 0$. On the other hand, a negative $\sqrt{\lambda}$ turns collapsing solutions into expanding solutions.}
\end{itemize}
            
Using the transformation equations (\ref{Phi}) and (\ref{T}), one can write the general solution for the scalar field $\phi$ as,
\begin{equation}
\label{phigen}
\phi\left( t\right) =\phi_{0}\left[ C^{\frac{1-n}{2}}\int^{t}f_{3}^{\frac{2}{n+3}%
}\left( \xi \right) e^{\left( \frac{1-n}{n+3}\right) \int^{\xi }f_{1}\left(
x \right) dx }d\xi -T_{0}\right] ^{\frac{2}{1-n}}f_{3}^{-\frac{1}{n+3}%
}\left( t\right) e^{-\frac{2}{n+3}\int^{t}f_{1}\left( x \right) dx
},
\end{equation}
where $\phi_{0}$ and $T_0$ are constants of integration and $C$ comes from the definition of the point transformations (\ref{Phi}) and (\ref{T}). Both $\phi_{0}$ and $C$ must be non-zero. Since the integrability criteria produces an exact time evolution of $A(t)$ as given in (\ref{A}), equation (\ref{phigen}) can be simplified in the present case as 
\begin{equation}
\label{phi-plw}
\phi\left( t\right) =\phi_{0}{{V_{0}}^{-\frac{1}{(n+3)}}}\Bigg(\frac{2n\sqrt{\lambda}}{n+3}\Bigg)^{-\frac{1}{n}}(t_{0}-t)^{-\frac{3}{n}}\Bigg[C^{\frac{1-n}{2}}{{V_{0}}^{\frac{2}{(n+3)}}}{\frac{n}{3}}\Bigg(\frac{2n\sqrt{\lambda}}{n+3}\Bigg)^{\frac{2}{n}}\Bigg((t_{0}-t)^{\frac{3}{n}}+\delta\Bigg)-T_0\Bigg]^{\frac{2}{(1-n)}},
\end{equation}

where $\delta$ comes as a constant of integration. One can clearly see that at $t=t_0$, when the volume element goes to zero, the scalar field diverges for $n>0$ and $n<-3$. A simple example for the evolution of the scalar field can be obtained where the integration constants $\delta$ and $T_0$ are put to zero. The time evolution then can be written as $\phi(t) \sim (t_0-t)^{\frac{3(1+n)}{n(1-n)}}$, which is consistent with the solution for scalar field one obtains from equation (\ref{wave-pwrlw}).         \\

From the field equations, one can write the expressions for the density and the pressure in terms of $A(t)$ and $\phi(t)$ as
\begin{equation}
\label{den-plw}
\rho=3kA^2+3\dot{A}^2-\frac{1}{2}A^2\dot{\phi}^2-\frac{{V_{0}}\phi^{n+1}}{n+1},
\end{equation}
\begin{equation}
\label{press-plw}
p=-kA^2+2\ddot{A}A-3\dot{A}^2-\frac{1}{2}A^2\dot{\phi}^2+\frac{{V_{0}}\phi^{n+1}}{n+1}.
\end{equation}

Both pressure and density diverge as $t$ goes to $t_0$ for $n > 0$ and $n < -3$. The expression for density indicates that if the scalar field part goes to infinity faster than the rest, the fluid density may go to a negative infinity close to the singularity. For a simple case, where $n=3$, i.e., the potential is defined as $V(\phi)=\frac{{V_{0}}\phi^4}{4}$, one can write $\phi = {\phi_0}{C_{1}}{\lambda}^{-\frac{1}{2}}(t_0-t)^{-2}$, for the arbitrary integration constants $T_0=\delta=0$ ($C_{1}$ is a constant depending on the values of $C$ and $V_{0}$). In this case, the scalar field part, contributing negatively, blows up much quicker ($\sim (t_0 - t)^{-8}$) as $t \rightarrow t_{0}$ than the rest, which go to infinity as  $\sim (t_0 - t)^{-2}$ and $\sim (t_0 - t)^{-4}$. However, the strong energy condition, $(\rho+3p) > 0$, can still be satisfied. From (\ref{den-plw}) and (\ref{press-plw}), one can write

\begin{equation}
(\rho+3p)=6\ddot{A}A-6\dot{A}^2-2A^2\dot{\phi}^2+\frac{2{V_{0}}\phi^{(n+1)}}{(n+1)}, 
\end{equation}

which can indeed remain positive. For the particular example of $n=3$, this can be simplified into
\begin{equation}
(\rho+3p) \sim \frac{6}{\lambda(t_0-t)^4}+\frac{{\phi_0}^2{C_{1}}^2}{2\lambda^2(t_0-t)^8}(V_{0}{\phi_0}^2{C_{1}}^2-16).
\end{equation}

Here, $\lambda$ is always positive as discussed earlier and so is ${\phi_0}^2{C_{1}}^2$. Near the singularity, as $t\rightarrow t_0$, the second term on the RHS becomes dominating over the first term. In order to satisfy the energy condition for all $'t'$, including the regime $t \rightarrow t_{0}$, the condition $V_{0}{\phi_0}^2{C_{1}}^2 > 16$ must be satisfied so that $(\rho+3p)$ remains positive definite. For $V_{0}{\phi_0}^2{C_{1}}^2 < 16$, $(\rho+3p)$ becomes negative when $t$ is close to the singular epoch $t_{0}$. It is intersting to note that in the Hawking radiation process, the stress energy tensor is known to behave in such a peculiar manner, such as the breakdown of weak energy condition $T^{\mu\nu}u_{\mu}u_{\nu} > 0$ in the classical sense, meaning a negative energy density \cite{roman}. It should also be noted that it is quite possible to ensure a positive $\rho$ by fixing the constants at the apparent horizon, which covers the singularity in all cases.

\subsection{The nature and visibility of the singularity}
The Kretschmann scalar can be calculated for the metric (\ref{metric}) as
\begin{equation}
\label{kretsch}
K=R_{\alpha\beta\gamma\delta}R^{\alpha\beta\gamma\delta}=6\dot{A}^4+6\Bigg(\frac{\ddot{A}}{A}-\frac{\dot{A}^2}{A^2}\Bigg)^2A^4,
\end{equation}
which, in view of the solution (\ref{Y}) yields \\

                               $$ K \sim (t_{0} - t)^{-4(n_{0}^{2} + 1)}$$, \\

for $n>0$ and $n<-3$. Since the kretschmann scalar clearly diverges as $t\rightarrow t_{0}$, one indeed has a curvature singularity as a result of the collapse.  \\

The standard analysis shows that the present singularity is a shell-focusing one (for which $g_{\theta\theta}=0$) and not a shell-crossing one (for which  $\frac{dg_{\theta\theta}}{dr}=0$, $g_{\theta\theta}\neq 0$ and $r>0$)\cite{yod, lake}.

Whether the curvature singularity is visible to an exterior observer or not, depends on the formation of an apparent horizon. The condition for such a surface is given by
\begin{equation}
\label{app-hor}
g^{\mu\nu}R,_{\mu}R,_{\nu}=0,
\end{equation}
where $R$ is the proper radius of the two-sphere, given by $\frac{r}{A(t)} = r Y(t)$ in the present case. The relevant cases in the present work are certainly the ones for $n > 0$ and $n < -3$. Using the explicit time evolution of $A$ from equation (\ref{A}), equation (\ref{app-hor}) yields a simple differential equation,
\begin{equation}
\label{app-hor1}
r^2\dot{Y}^2-(1-kr^2)Y^2=0,
\end{equation}
which, in view of equations (\ref{Y}) and (\ref{scale1}), yields the algebraic equation at $t=t_{app}$
\begin{equation}
\label{app-hor2}
\frac{\dot{Y}}{Y} = \frac{n_{0}^{2}}{t_{app}-t_{0}}.
\end{equation}

Since the present interest is in a collapsing solution, scale factor must be a monotonically decreasing function of time. So $\dot{Y}$ is negative and $Y$, being the scale factor, must always be positive. Thus from equation (\ref{app-hor2}), the condition is consistent if and only if $t_{app} < t_{0}$. This clearly indicates that the apparent horizon forms before the formation of the singularity, for all relevant cases. Thus, the curvature singularity is always covered from an exterior observer by the formation of an apparent horizon. At the singularity in the present case, one has $Y = 0$ and $\dot{Y} \neq 0$. Equation (\ref{app-hor1}) indicates that this is consistent only with $r = 0$ at the singularity. Thus the singularity is strictly a central singularity which could have been a naked singularity as well, as discussed by Christodoulou \cite{christozero}. It deserves mention that had the singularity been independent of the radial coordinate $r$, it would have been certainly covered by a horizon, discussed by Joshi, Goswami and Dadhich \cite{naresh}.

\subsection{Matching with an exterior Vaidya spacetime}
Generally, in collapsing models, a spherically symmetric interior is matched with a suitable exterior solution, a Vaidya metric or a Schwarzschild metric depending on the prevailing conditions \cite{dwivedi}. This requires the continuity of both the metric and the extrinsic curvature on the boundary hypersurface. We choose the radiating Vaidya solution as a relevant exterior to be matched with the collapsing interior, defined as
\begin{equation}
\label{metric2}
ds^2=\frac{1}{{A(t)}^2}\Bigg[dt^2-dr^2-r^2d\Omega^2\Bigg].
\end{equation}
Vaidya metric is given by
\begin{equation}\label{vaidya}
ds^2=\Bigg[1-\frac{2m(v)}{R}\Bigg]dv^2+2dvdR-R^2d\Omega^2.
\end{equation}
The quantity $m(v)$ represents the Newtonian mass of the gravitating body as measured by an observer at infinity. The metric (\ref{vaidya}) is the unique spherically symmetric solution of the Einstein field equations for radiation in the form of a null fluid.  The necessary conditions for the smooth matching of the interior spacetime to the exterior spacetime was presented by Santos \cite{santos} and also discussed in detail by Chan \cite{chan}, Maharaj and Govender \cite{goven} in context of a radiating gravitational collapse. Following their work, the relevant equations matching (\ref{metric2}) with (\ref{vaidya}) can be written as                         
\begin{equation}
\Bigg[\frac{r}{A(t)}\Bigg]_{\Sigma} = R,
\end{equation}

\begin{equation}
\Bigg[r(r B')\Bigg]_{\Sigma} = \Bigg[R A \Big(1-\frac{2m(v)}{R}\Big)\dot{v} + RA\dot{R}\Bigg]
\end{equation}

\begin{equation}
m(v)_{\Sigma} = \frac{r^3}{2A^3}\Bigg(\dot{A}^2-A'^2+\frac{A'A}{r}\Bigg),
\end{equation}
and
\begin{equation}\label{prq}
p_{r} = \Bigg[\frac{q}{A(t)}\Bigg]_{\Sigma} = 0,
\end{equation}                        
where $\Sigma$ is the boundary of the collapsing fluid and $q$ denotes any radial heat flux defined in the interior of the collapsing scalar field.     \\
Equation (\ref{prq}) yields a nonlinear differential condition between the conformal factor and the scalar field to be satisfied on the boundary hypersurface $\Sigma$ as
\begin{equation}\label{bcondition}
\Bigg[2\frac{\ddot{A}}{A} - 2\frac{\dot{A}^2}{A^2} - \frac{1}{2}\dot{\phi}^2 + \frac{V_{0}}{A^2}\frac{\phi^{(m+1)}}{(m+1)}\Bigg]_{\Sigma} = 0.
\end{equation}
Using the time evolution of the conformal factor and the scalar field, i.e. equations (\ref{A}) and (\ref{phi-plw}), one can simplify this expression and establish some restrictions connecting the parameters such as $V_{0}$, $n$, $\lambda$, $\phi_{0}$, $\delta$ and $T_{0}$. Therefore, the validity of the present model is established alongwith certain constraints.   \\

An interesting feature is observed if the interior solution is matched with a Schwarzschild exterior. On the boundary hypersurface $\Sigma$, the matching of extrinsic curvature gives
\begin{equation}
\Bigg[\frac{2{n_{0}}^2-{n_{0}}^4}{(t-t_0)^2}\Bigg]_{\Sigma} = 0,
\end{equation}
which means ${n_{0}}^{2} = 2$ and it is easy to note that the resulting metric corresponds to the Oppenheimer-Snyder model for the marginally bound case.  \\

However, it must be noted that $(\frac{n+3}{2n}) = n_{0}^2 = 2$ implies that $n = 1$, which does not fall in the domain of validity of the theorem employed in this work.

\section{Potential as a combination of the form $V(\phi) = \frac{1}{2}{\phi}^{2} + \frac{\phi^{n+1}}{n+1}$}

For a very simple combination of two powers of $\phi$,
\begin{equation}
 \label{poly}
V(\phi) = \frac{1}{2}{\phi}^{2} + \frac{\phi^{n+1}}{n+1},
\end{equation}
the method of integrability of anharmonic oscillators can lead to some interesting informations about the behaviour of the collapse. With equation (\ref{poly}), one can write 
\begin{equation}
\label{poly3}
\frac{dV}{d\phi}=\phi+\phi^n.
\end{equation}

The scalar field equation (\ref{wave}), with the same metric (\ref{metric}), becomes
\begin{equation}
\label{wave-poly}
\ddot{\phi}-2\frac{\dot{A}}{A}\dot{\phi}+\frac{\phi}{A^2}+\frac{\phi^n}{A^2}=0,
\end{equation}
which can be written in a general form 
\begin{equation}
\label{wave-poly1}
\ddot{\phi}+f_1(t)\dot{\phi}+f_2(t)\phi+f_3(t)\phi^n=0,
\end{equation}
in a similar way as in the case of a simple power law potential. It is easy to recognize $f_{i}$'s as $f_{1} = -2\frac{\dot{A}}{A}, f_{2} = f_{3} = \frac{1}{A^{2}}$. Equation (\ref{int-gen}) now reduces to

\begin{equation}
\label{wave-poly2}
\frac{\ddot{A}}{A}-3\frac{(n+1)}{(n+3)}\frac{\dot{A}^2}{A^2}+\frac{(n+3)}{6A^2}=0.
\end{equation}

This differential equation yields a straightforward first integral given by
\begin{equation}
\dot{A}^2-{\lambda}A^{\frac{6(n+1)}{(n+3)}}-\frac{(n+3)^2}{18(n+1)}=0,
\end{equation}
where $\lambda$ comes as a constant of integration. This can be written in a simpler form,
\begin{equation}\label{wave-poly3}
\dot{A}=\Big({\lambda}A^p+q\Big)^\frac{1}{2},
\end{equation}
where, $p=\frac{6(n+1)}{(n+3)}$ and $q=\frac{(n+3)^2}{18(n+1)}$.
\par The general solution of equation (\ref{wave-poly3}) can in fact be given in the form of Gauss' Hypergeometric function,

\begin{equation}
\label{sol-A-poly}
\frac{A}{\sqrt{q}}{_2}F{_1}\Bigg[\frac{1}{2},\frac{1}{p};(1+\frac{1}{p});-\frac{\lambda A^p}{q}\Bigg]=t-t_0,
\end{equation}
where $t_{0}$ is a constant of integration. \\

It is very difficult to invert the equation (\ref{sol-A-poly}) and write $A(t)$ as a function of $t$ explicitly. However, since we are interested in a regime of space-time, where the volume is very small, an approximate analysis of this equation can be given, assuming $A(t)\rightarrow\infty$, meaning the proper radius ($\sim \frac{1}{A}$) is very small. From the series expansion of the Hypergeometric function, one can write 
\begin{equation}
\label{large-x}
{_2}F_{1}(a,b;c;x)=\frac{\Gamma(b-a)\Gamma(c)}{\Gamma(b)\Gamma(c-a)}(-x)^{-a}\Bigg[1+\textit{O}\Bigg(\frac{1}{x}\Bigg)\Bigg]+\frac{\Gamma(a-b)\Gamma(c)}{\Gamma(a)\Gamma(c-b)}(-x)^{-b}\Bigg[1+\textit{O}\Bigg(\frac{1}{x}\Bigg)\Bigg]
\end{equation}
for $|x|\rightarrow\infty$, $a\neq b$.  \\

Using (\ref{large-x}), the expression for $A$ can be written from equation (\ref{sol-A-poly}) as
\begin{equation}
\label{large-A}
\frac{A^{1-p/2}}{\lambda^{1/2}}\frac{\Gamma(1/p-1/2)\Gamma(1+1/p)}{\Gamma(1/p)\Gamma(1/p+1/2)}\Bigg[1+\textit{O}\Bigg(-\frac{q}{\lambda A^p}\Bigg)\Bigg]+\frac{q^{1/p-1/2}}{\lambda^{1/p}}\frac{\Gamma(1/2-1/p)\Gamma(1+1/p)}{\Gamma(1/2)\Gamma(1)}\Bigg[1+\textit{O}\Bigg(-\frac{q}{\lambda A^p}\Bigg)\Bigg]=t-t_0.
\end{equation}

\begin{figure}[h]
\begin{center}
\includegraphics[width=0.5\textwidth]{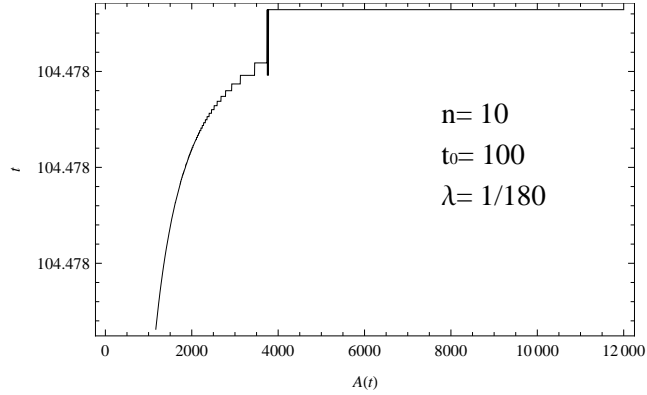}
\caption{Plot of $t$ vs $A(t)$ for $n=10$ and a positive $\lambda$.}
\end{center}
\label{fig:hyp1}
\end{figure}

A careful study of equation (\ref{large-A}) reveals that, for all $(1-\frac{p}{2})<0$, $A(t)\rightarrow\infty$, which implies that the scale factor ${\frac{1}{A(t)}\rightarrow 0}$, for a negative $(1-\frac{p}{2})$, at the time 
\begin{equation}
t=t_0+\Bigg[\frac{q^{1/p-1/2}}{\lambda^{1/p}}\Bigg]\Bigg[\frac{\Gamma(1/2-1/p)\Gamma(1+1/p)}{\Gamma(1/2)\Gamma(1)}\Bigg].
\end{equation}
It should be noted that as this would require $(1-\frac{p}{2})$ to have a negative value, $n$ is either positive or $n<-3$. The latter however will lead to imaginary solutions for the scale factor and will not be considered in the subsequent discussion.

\begin{figure}[h]
\begin{center}
\includegraphics[width=0.5\textwidth]{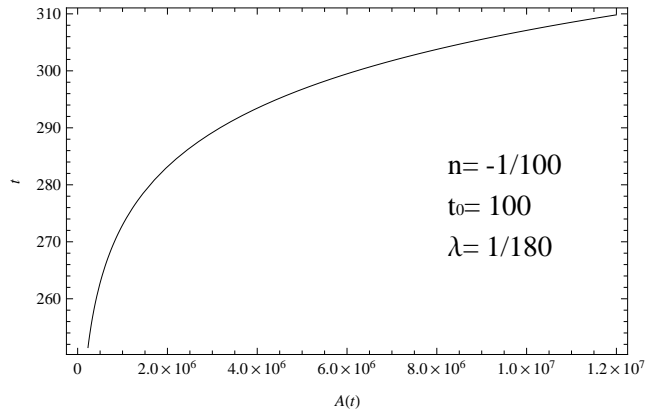}
\caption{Plot of $t$ vs $A(t)$ for $n=-\frac{1}{100}$ and a positive $\lambda$.}
\end{center}
\label{fig:hyp1}
\end{figure}

For the sake of completeness, we should mention that the general solution for the scalar field equation can be written as 
\begin{equation}
T=T_{0}+\frac{\epsilon }{C_{0}}\Phi\sqrt{\frac{C_{0}\left( n+1\right) -{\Phi}^{n+1}}{%
2\left( n+1\right) }}\,_{2}F_{1}\left[ 1,\frac{n+3}{2\left( n+1\right) };%
\frac{n+2}{n+1};\frac{{\Phi}^{n+1}}{C_{0}(n+1)}\right] , \qquad   n\neq -1,
\end{equation}
where $T_0$ and $C_0$ are arbitrary constants of integration and  $\Phi$ and $T$ are defined by equations (\ref{Phi}) and (\ref{T}) respectively.

\begin{figure}[h]
\begin{center}
\includegraphics[width=0.5\textwidth]{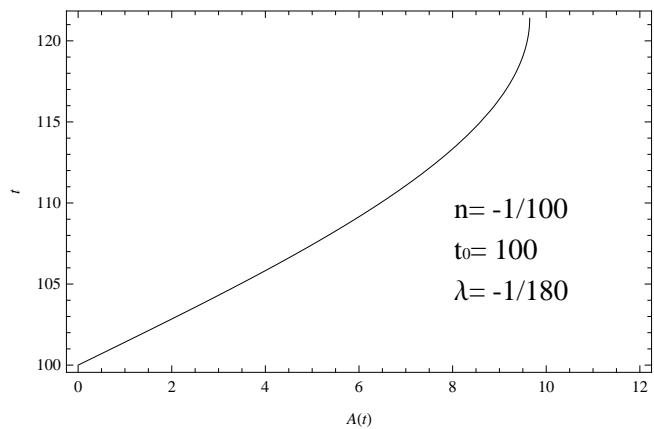}
\caption{Plot of $t$ vs $A(t)$ for negative $\lambda$.}
\end{center}
\label{fig:hyp1}
\end{figure}

We shall discuss a few examples with some values of the constants $n, \lambda$ and $t_{0}$ with the help of numerical plots. Figure $1$ shows that for $n=10$ and a positive $\lambda$, A increases very fast to an indefinitely large value at a finite value of time $t$ indicating that the proper radius ($\frac{1}{A}$) and hence the proper volume indeed crushes to a singularity. Figure $2$ shows that for a small negative value of $n$, namely $n=-\frac{1}{100}$, one has a collapsing situation, but the rate of collapse dies down and the singularity is not reached at a finite time.  This is quite consistent with the inference drawn from equation (\ref{large-A}) that for a collapsing sphere to reach a singular state at a finite time, one would require a positive value of $n$. The behaviour is also sensitive to the initial conditions. For example, for a negative $\lambda$, the same small negative value of $n$ would lead to a situation where the distribution will not collapse beyond a certain constant finite volume at a finite time, as shown in figure $3$.

\section{Discussion}
A spherically symmetric scalar field collapse is investigated in the present work. Only a limited amount of work on a massive scalar field collapse can be found in the literature and that too in a very restricted scenario. In the present work we discuss a massive scalar field collapse with a power law potential ($V \sim {\phi}^{n+1}$) in a very general situation which includes a very wide range of the values of $n$. \\

In order to study the problem analytically, we adopt a strategy of dealing with the integrability condition for the scalar field equation. The recently developed technique of solving anharmonic oscillator problem by Euler \cite{euler} (see also \cite{harko}) has been utilized. \\

It is interesting to note that the conclusions drawn from these calculations are independent of the choice of any equation of state for the fluid distribution. This is because the scale factor is calculated straightaway from the integrability condition. The field equations can be utilized in the determination of the fluid density and pressure as functions of $A$ and $\phi$ and hence as a function of $t$ (equations (\ref{den-plw}) and (\ref{press-plw})).  \\

The general result is that it is indeed possible to have a collapsing situation which crushes to the singularity of zero proper volume and infinite curvature. This situation is observed for potentials of the form $V(\phi) \sim {\phi}^{n+1}$ where $n < -3$ or $n > 0$. However, for $0 > n > -3$, the distribution collapses for ever, reaching the singularity only at an infinite future. \\

We find that for a continuous gravitational collapse of a massive scalar field with potential of the form $V(\phi) \sim \phi^{(n+1)}$, whenever one has a singularity at a finite future, it is necessarily covered by a horizon. This is completely consistent with the theorem proved by Hamid, Goswami and Maharaj, that for a continuous gravitational collapse in a conformally flat spacetime, the end product is necessarily a black hole \cite{hgm}.      \\

A quadratic potential is of a primary interest in scalar field theories. But this form of potential is out of the domain of validity of the theorem used (the method does not work for $n=1$, which corresponds to a quadratic potential). In the fifth section, we include a discussion on a potential containing two terms one of which is a quadratic in ${\phi}$. Although an elaborate discussion like a simple power law have not been possible, quite a few interesting results from the asymptotic behaviour of the solution has been noted with the help of numerical plots. Depending on the initial conditions, there are many interesting possibilities, where the singularity is reached only at infinite time, and even a situation where the collapsing object settles down to a finite size rather than crushing into a singular state. The last scenario, illustrated in figure $3$, looks like a white dwarf or a neutron star where the collapsing star equilibriates as a finite object when the degenerate fermion pressure is able to halt the gravitational collapse. Apparently a scalar field with a potential which is a power law of the field $\phi$ with a small negative exponent along with a ${\phi}^{2}$ term can also do the trick. \\

The present work adds to the existing literature in two ways. First, it fills in for the paucity of information in connection with massive scalar field collapse. Second, it shows the usefulness of Euler's Theorem on the integrability of anharmonic oscillator equation in mining information from apparently hopeless situations in the context of scalar field collapse.

\end{document}